\documentclass[conference]{IEEEtran}
\IEEEoverridecommandlockouts
\usepackage{cite}
\usepackage{amsmath,amssymb,amsfonts}
\usepackage{algorithmic}
\usepackage{graphicx}
\usepackage{textcomp}
\usepackage{xcolor}
\usepackage{capt-of}  
\usepackage{cuted}    
\usepackage{multirow}
\usepackage{nicematrix}
\usepackage{amsmath}
\DeclareMathOperator*{\argmax}{arg\,max}

\usepackage[linesnumbered,ruled,vlined]{algorithm2e}

\newcommand{\ket}[1]{\ensuremath{\left|#1\right\rangle}}

\usepackage{braket}
\usepackage{lipsum}
\def\BibTeX{{\rm B\kern-.05em{\sc i\kern-.025em b}\kern-.08em
    T\kern-.1667em\lower.7ex\hbox{E}\kern-.125emX}}
\begin{document}

\title{A Comparison of Encoding Techniques for an Analog Quantum Emulation Device\\}

\author{\IEEEauthorblockN{ Sharan Mourya}
\IEEEauthorblockA{\textit{Indian Institute of Technology, Hyderabad} \\
sharan.mourya@5g.iith.ac.in}}

\maketitle

\begin{abstract}
Quantum computers can outperform classical computers in certain tasks. However, there are still many challenges to the current quantum computers such as decoherence and fault tolerance, and other drawbacks such as portability and accessibility. In this study, we circumvent these issues by realizing an analog quantum emulation device (AQED) where each qubit state is represented by a unique analog signal. It is possible to do this because previously it was shown that Hermitian operations on a Hilbert space are not unique to quantum systems and can also be applied to a basis of complex signals that form a Hilbert space. Orthogonality of the complex signal basis can be maintained by separating the signals into the frequency domain or the spatial domain. We study both these approaches and present a comparison. We finally realize the entire device on a UMC 180nm processing node and demonstrate the computational advantage of an AQED by emulating Grover's search algorithm (GSA) and Quantum Fourier Transform (QFT). We also present the equivalent quantum volume achieved by this device. 
\end{abstract}


\section{Introduction}
A quantum computer provides a new computational paradigm that is fundamentally different from a classical computer\cite{feynman}. This revolutionary idea reconciles the theory of computation with quantum physics and has shown great promise in solving a broad variety of problems that are originally not solvable on traditional computers \cite{zoller}. Following Moore's law, classical computers are getting increasingly dense and  ever so slightly approaching the quantum domain which makes quantum computers the ideal candidate to replace classical ones as they fundamentally exploit the quantum phenomena. Quantum computers have also shown tremendous speed up in certain operations compared to classical ones. Google has shown this quantum supremacy in its seminal paper \cite{google}.\par
There are various ways to realize a quantum computer such as using superconducting circuits \cite{super}, trapped ions \cite{trap}, photons, and neutral atoms \cite{neutral}. In all these approaches maintaining coherence between the states is a difficult task as the slightest of interaction with the surroundings will cause the system to lose its coherence and perform poorly. To overcome these difficulties a novel way of designing a universal quantum computer based on classical signals was introduced in \cite{main}. A qubit may be represented as a pair of complex signals each corresponding to a binary state. Superposition can then be considered as a scaled sum of these two complex signals and the orthogonality of these signals can be maintained by choosing the signals as in-phase and quadrature-phase signals of a Quadrature Amplitude Modulation (QAM) encoding scheme. A composite qubit system can then be constructed by utilizing more complex signals to represent the states. This approach uses a classical, signal-based model to represent a multi-qubit quantum system along with all the gate operations and also allows us to apply a gate operation on a single qubit or multiple qubits, or all qubits at once. There are a lot of advantages to this approach such as no decoherence, no need to operate the device close to absolute zero, and less complexity of the circuit. Another non-trivial advantage of using an emulator is using this to solve problems in $\textbf{NP}$ and $\textbf{\#P}$ as suggested by Abrams and Lloyd by adding non-linearity to the operators. Unfortunately, quantum mechanics has strictly linear time evolution but an emulation device operating on signals is not limited by that\cite{adv}. Even though this is a very promising approach that exhibits inherent parallelism as much as any other quantum computer and can take advantage of quantum entanglement, this approach is not scalable \cite{main}. \par
To address the issue of scalability in this model, another approach was introduced \cite{parallel} whereby repeating the qubit complex signal either in the spatial domain or temporal domain increases the total number of qubits in the system. This solves the issue of scalability to some extent but the size of the circuit still grows exponentially with the number of qubits which makes this device to be used only in a compact form with limited qubits. This can be used as a co-processor performing accelerated tasks besides a digital processor similar to the FPGA-based accelerators used in deep learning \cite{acc}. In this work, we would like to design an analog quantum emulation device (AQED) that uses the repetition of qubit signals in the spatial domain to increase the number of qubits. We chose to design a $6$ qubit system emulating two of the most important quantum algorithms: Grover's Search Algorithm (GSA) \cite{gsa} and Quantum Fourier Transform (QFT) \cite{qft}. \par
The organization of this paper is as follows. In Section II we summarize the frequency-based encoding followed by presenting the system-level implementation of this in Section III. In Section IV, we discuss the spatial encoding method and its advantages over spectral encoding. Section V deals with the implementation of Grover's search algorithm followed by the Quantum Fourier transform in Section VI. Then we present the results in Section VII.

\section{Spectral Encoding}
Quantum physics utilizes the mathematical structure of Hilbert space with composite systems defined as a tensor product of Hilbert spaces. This structure is not unique to quantum systems. We can use complex exponential signals to represent such a system. \par
Consider a single qubit be represented by the two quadrature signals $e^{j\omega_{0} t}$ and $e^{-j\omega_{0} t}$ over some interval $t \in [0,T)$ corresponding to the states $\ket{0}$ and $\ket{1}$ respectively. With this, any state $\ket{\psi}=a\ket{0}+b\ket{1}$ can be represented as $\ket{\psi} = ae^{j\omega_{0}t}+be^{-j\omega_{0}t}$ where $a,b \in \mathbb{C}$. Note that, we can represent any superposed state uniquely with this representation. To finish constructing the Hilbert space, we define the inner product of any two states $\ket{\phi}$ and $\ket{\psi}$ as follows
\begin{equation}
    \braket{\varphi|\psi} = \frac{1}{T} \int_{0}^{T}\varphi(t)^{*}\psi(t) \,dt
    \label{inner}
\end{equation}
From this we can calculate that $\braket{0|0} = \braket{1|1} = 1$ and $\braket{0|1} = \braket{1|0} = 0$. This completes a single qubit representation and to further add more qubits we use in-phase and quadrature signals of different frequencies. In particular, for $n$ qubits we use $\omega_{0} < \omega_{1} < \omega_{2} \cdots < \omega_{n-1}$. We define $n$ different single qubit systems as described above with $n$ different frequencies and multiply all of them to represent an $n$ qubit system. This final signal will contain $2^n$ frequencies, half of which will be negative. To distinctly represent these $2^n$ frequencies we use octave spacing $w_{i} = 2^i\omega_{0}$. With this, all possible sums and differences of frequencies will be unique and uniformly spaced. Each such product forms one of the $2^n$ basis signals. For example, $\ket{\phi_{x}}$, for $x=\sum_{i=0}^{n-1}x_{i}2^i$ where $x_{i} \in \{0,1\}$, can be represented as 
\begin{equation}
    \ket{\phi_{x}} = exp\bigg[\sum_{i=0}^{n-1}(-1)^x_{i}j\omega_{i}t\bigg]
    \label{basis}
\end{equation}
Following this, an $n$-qubit system can be represented by a linear combination of $N=2^n$ basis states.
\begin{equation}
    \ket{\psi} = \sum_{x=0}^{N}a_{x}\ket{\phi_{x}}
    \label{sum_basis}
\end{equation}
where $a_{x} \in \mathbb{C}$. Note that, any $n$ qubit system can be uniquely represented by this.

\subsection{Gate Operation}
We have obtained the complex signal representation of an $n$ qubit system which is a linear combination of $N$ basis states. To apply a gate, we need to obtain the amplitudes of each basis state in the superposition, i.e., $a_{x}$. We can obtain this using the definition of inner product as follows
\begin{equation}
    \braket{\phi_{x}|\psi} = \sum_{y=0}^{N}a_{y}\braket{\phi_{x}|\phi_{y}} = a_{x}
    \label{decomp}
\end{equation}
This inner product is simply a sub-space projection of $\ket{\psi}$. We generalize this by defining a projection operation as follows
\begin{equation}
    \Pi_{a}^{(i)}\ket{\psi} = \ket{a}_{i} \otimes \ket{\psi_{a}^{i}}
    \label{projpi}
\end{equation}
$\Pi_{a}^{i}$ is defined as the projection of $\ket{\psi}$ on the $i^{th}$ basis state. This $i^{th}$ basis state could be either $\ket{0}$ or $\ket{1}$ which is captured by $a$ that could be either $0$ or $1$. So for example,
\begin{equation}
    \Pi_{0}^{(i)}\ket{\psi} = \ket{0}^{(i)} \otimes \ket{\psi_{0}^{i}} = \sum_{x:x_{i}=0}a_{x}\ket{x_{n-1} x_{n-2} \cdots 0 \cdots x_{1} x_{0}}
    \label{pi0}
\end{equation}
\begin{equation}
    \Pi_{1}^{(i)}\ket{\psi} = \ket{1}_{i} \otimes \ket{\psi_{1}^{i}} = \sum_{x:x_{i}=1}a_{x}\ket{x_{n-1} x_{n-2} \cdots 1 \cdots x_{1} x_{0}}
    \label{pi1}
\end{equation}
From equations \eqref{pi0} and \eqref{pi1} we can write
\begin{equation}
    \ket{\psi} = \Pi_{0}^{(i)}\ket{\psi} + \Pi_{1}^{(i)}\ket{\psi}
    \label{pi0+pi1}
\end{equation}
Let us consider a 2-qubit system
\begin{equation}
    \ket{\psi} = a_{0}e^{3j\omega_{0}t}+a_{1}e^{j\omega_{0}t}+a_{2}e^{-j\omega_{0}t}+a_{3}e^{-3j\omega_{0}t}
    \label{2qubit}
\end{equation}
The projections for this system are
\begin{equation}
    \Pi_{0}^{(i)}\ket{\psi} = e^{2j\omega_{0}t}(a_{0}e^{j\omega_{0}t}+a_{1}e^{-j\omega_{0}t})
    \label{pi0_2}
\end{equation}
\begin{equation}
    \Pi_{1}^{(i)}\ket{\psi} = e^{-2j\omega_{0}t}(a_{2}e^{j\omega_{0}t}+a_{3}e^{-j\omega_{0}t})
    \label{pi1_2}
\end{equation}
Equation \eqref{pi0+pi1} can be verified by adding equations (10) and (11). Physically projection operation is implemented as a complex multiplication followed by a filtering operation as discussed in \cite{main}. \par
Combining equations \eqref{pi0}, \eqref{pi1} and \eqref{pi0+pi1} we can write $\ket{\psi}$ as
\begin{equation}
    \ket{\psi} = \ket{0}^{(i)} \otimes \ket{\psi_{0}^{i}} + \ket{1}^{(i)} \otimes \ket{\psi_{1}^{i}}
    \label{pi0_2+pi1_2}
\end{equation}
With this we can define a single qubit gate operation as
\begin{equation}
    U^{(i)}\ket{\psi} = \big(U\ket{0}^{(i)}\big) \otimes \ket{\psi_{0}^{i}} + \big(U\ket{1}^{(i)}\big) \otimes \ket{\psi_{1}^{i}}
    \label{upsi}
\end{equation}
where $U^{(i)}$ is the gate $U$ acting on the $i^{th}$ qubit and 
\begin{equation}
    U\ket{0} = U_{00}\ket{0} + U_{10}\ket{1}
    \label{u0}
\end{equation}
\begin{equation}
    U\ket{1} = U_{01}\ket{0} + U_{11}\ket{1}
    \label{u1}
\end{equation}
\begin{equation}
U = 
\begin{pmatrix}
    U_{00} & U_{01}\\
    U_{10} & U_{11}
\end{pmatrix}
\label{umat}
\end{equation}

With this, the new state looks like
\begin{align}
\begin{split}
    \psi'(t) = \big(U_{00}e^{j\omega_{i}t}+U_{10}e^{-j\omega_{i}t}\big)\psi_{0}^{(i)}(t)\\ + \big(U_{01}e^{j\omega_{i}t}+U_{11}e^{-j\omega_{i}t}\big)\psi_{1}^{(i)}(t)
\end{split}
\label{UPsi}
\end{align}

\section{System Model}
\subsection{Single Qubit}
\subsubsection{Projection}
A one qubit state can be represented by $\psi(t) = a_{0}e^{j\omega t} + a_{1}e^{-j\omega t}$ where $\ket{0} = e^{j\omega t}$ and $\ket{1} = e^{-j\omega t}$ are the two basis states. Projection of the state $\ket{\psi}$ on any of the basis can be achieved by performing the inner product in Equation 4 which is a multiplication followed by a filtering operation. Projection on $\ket{0}$ is as follows
\begin{equation}
    e^{-j\omega t}\psi(t) = a_{0} + a_{1}e^{-2j\omega t}
    \label{decomp0}
\end{equation}
after filtering by a low pass filter with a cut-off frequency of $\omega$, we get 
\begin{figure*}[h]
\centering
\includegraphics[width=6in,height=3.5in]{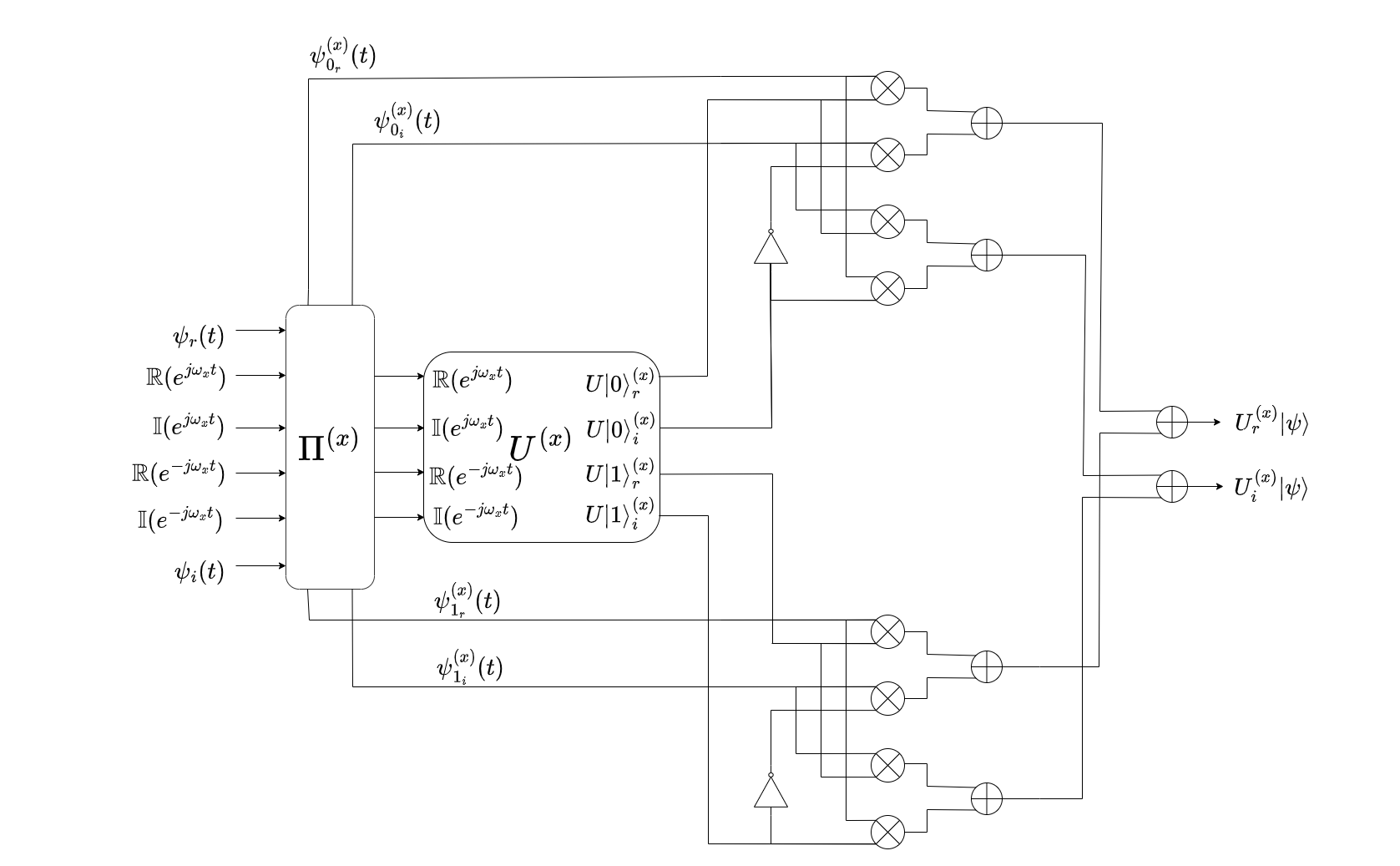}
\captionof{figure}{One Qubit Gate Operation}
\label{one_gate}
\end{figure*}
\begin{equation}
     \braket{0|\psi} = a_{0}
     \label{decompa0}
\end{equation}
Similarly for a projection on $\ket{1}$ we get,
\begin{equation}
     \braket{1|\psi} = a_{1}
     \label{decompa1}
\end{equation}
Note that, $\ket{\psi_{0}^{0}} = a_{0}$ and $\ket{\psi_{1}^{0}} = a_{1}$.

\begin{figure}
\centering
\includegraphics[width=3in,height=1.7in]{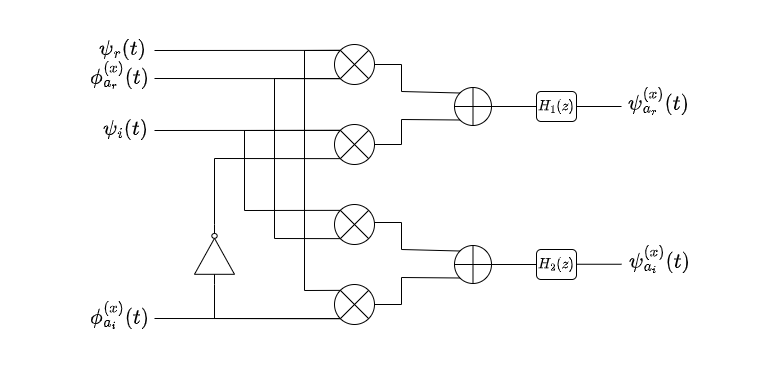}
\captionof{figure}{Projection}
\label{proj}
\end{figure}

Also, the multiplication in Eq \eqref{decomp0} is complex so physically it consists of real and imaginary parts multiplying separately as shown in Fig.{~\ref{proj}}. This complex multiplication is followed by two null-space rejection filters (as the filter removes the null space of the operator) one for the real part and the other for the imaginary part. 
In Fig.{~\ref{proj}} $\psi_{r}(t)$ and $\psi_{i}(t)$ are real and imaginary parts of $\psi(t)$.  $\phi_{a_{r}}^{(x)}$ and $\phi_{a_{i}}^{(x)}$ are the real and imaginary parts of $\phi_{a}^{(x)}$ which is the basis function of qubit $x$ which is in the state $a \in \{0,1\}$. Thus Fig.{~\ref{proj}} represents the projection of $\psi(t)$ on $\phi_{a}^{(x)}$.

\subsubsection{Gate Operation}
\begin{figure}[h]
\centering
\includegraphics[width=3.5in,height=2.7in]{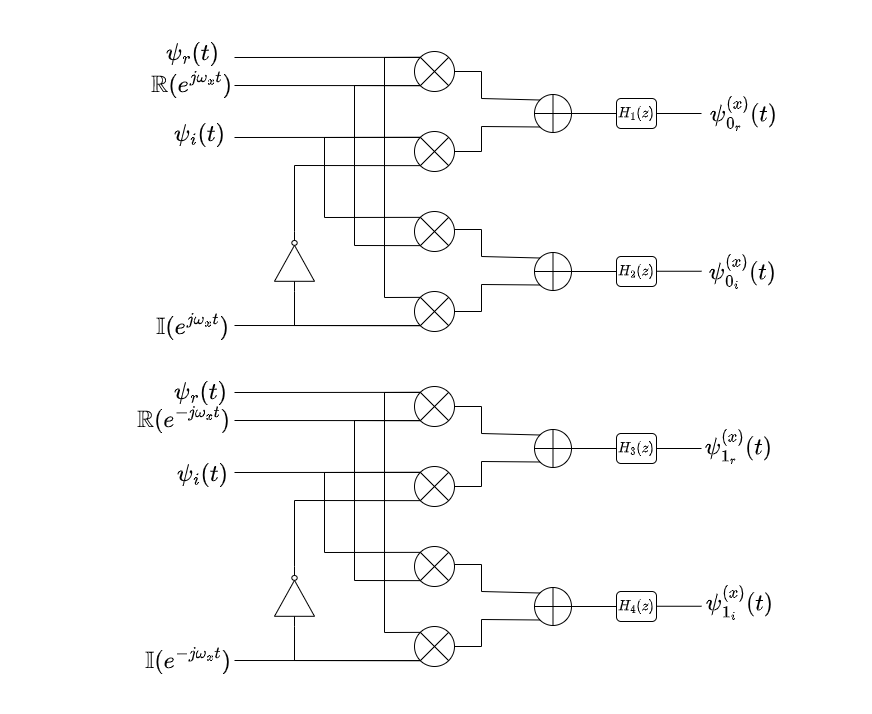}
\captionof{figure}{Projection Block}
\label{proj_block}
\end{figure}

For a single qubit gate operation, we need the projections of state $\psi(t)$ on both the bases $\ket{0}$ and $\ket{1}$. So, we need two such projection blocks one for each basis followed by the gate operation as described in Eq \eqref{UPsi} which is shown in Fig.{~\ref{proj_block}}.
Also, from Eq (12) and Fig.{~\ref{proj_block}}, note that $\ket{\psi_{0}^{(x)}} = \ket{\psi_{0_{r}}^{(x)}} + j\ket{\psi_{0_{i}}^{(x)}}$ and $\ket{\psi_{1}^{(x)}} = \ket{\psi_{1_{r}}^{(x)}} + j\ket{\psi_{1_{i}}^{(x)}}$. Similarly, $\ket{0}^{(x)} = e^{j\omega_{x} t}$ and $\ket{1}^{(x)} = e^{-j\omega_{x} t}$. Now with the projections available, we can apply the gate operation as shown in Fig.{~\ref{one_gate}}.\par

$\Pi^{(x)}$ is the projection block as shown in Fig 2 and $U^{(x)}$ is the gate operation as described in Eq \eqref{upsi}. Fig.{~\ref{one_gate}} essentially captures how a qubit is projected onto a subspace and operates it with a single qubit gate and then combines both to get the evolved state.
\subsubsection{Implementation}
\begin{figure}[h!]
\centering
\includegraphics[width=3.5in,height=0.7in]{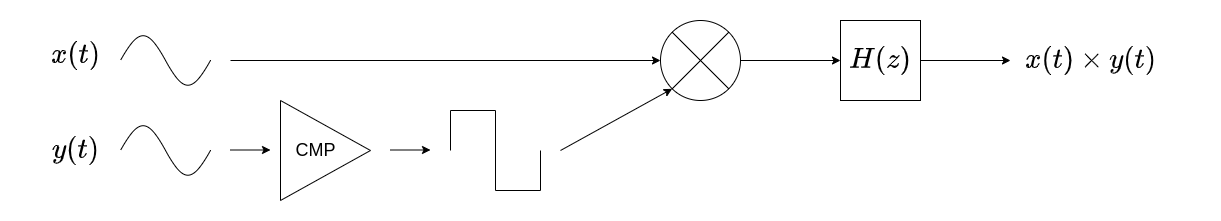}
\captionof{figure}{Multiplication}
\label{multiplication}
\end{figure}

To implement a full single qubit gate, we need a multiplier circuit, an adder, and an inverter. Adder and inverter are straightforward to implement using opamps which leaves the multiplier as the bottleneck. Gilbert cell multiplier or a four-quadrant multiplier can be used here but they consume a considerable area given the number of multipliers required while scaling the number of qubits. To circumvent this, we modify the mixer operation as shown in Fig.{~\ref{multiplication}}.

\begin{figure}[h!]
\centering
\includegraphics[width=2in,height=1in]{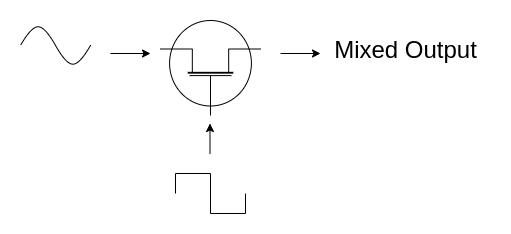}
\captionof{figure}{Mixer}
\label{mixer}
\end{figure}

Here, instead of directly multiplying the two sinusoids, we first convert one of them into a square wave using an opamp comparator. We then multiply the two signals and use a low pass  filter to reject the additional harmonics of the square wave. This might look like a complicated way to multiply two signals but this requires less area than that of direct multiplication as in this case we can simply replace the multiplier with a MOS switch whose gate is driven by the square wave and source terminal is driven by the sine wave. This is shown in Fig.{~\ref{mixer}}.\par
\begin{figure}[h]
\centering
\includegraphics[width=3in,height=0.85in]{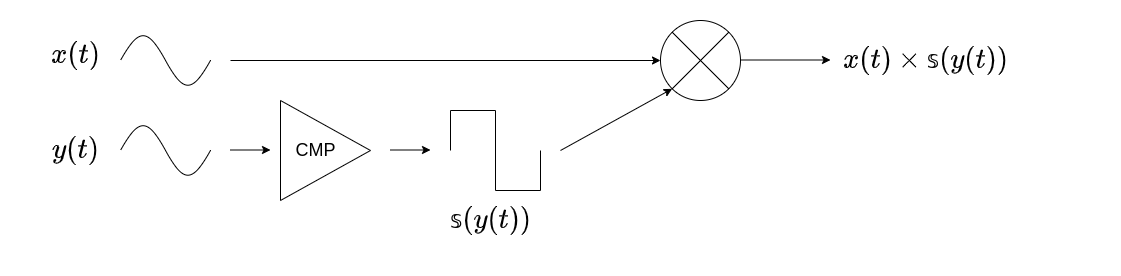}
\captionof{figure}{Simplified Multiplication}
\label{smixer}
\end{figure}

Furthermore, for a single qubit system we deal with only one frequency, say $\omega$. Then we need a low pass filter with a cut-off frequency of $2\omega$ to remove all the harmonics. But, in Fig.{~\ref{proj_block}}, there's already a bank of filters at the end of the projection operation and for a single qubit system, these filters are all identical with a cut-off frequency of $\omega$. So even if we don't filter out the harmonics right after mixing, they will be filtered out at the end of the projection. Hence, our multiplication operation reduces to Fig.{~\ref{smixer}}.\par

\begin{figure}[h]
\centering
\includegraphics[width=3.7in,height=2.2in]{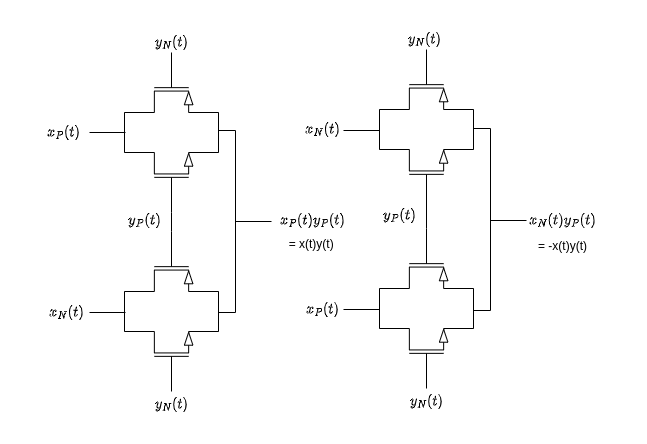}
\captionof{figure}{Transmission gate mixer}
\label{tmixer}
\end{figure}

In the mixer, we used a transmission gate as the MOS switch. Also, we made this a differential design to make it noise-resilient. With this design methodology, we don't need an inverter as we can just swap the two differential signals and for the adder, we used a single-stage opamp adder. Schematics of adders and filters are omitted in this paper for simplicity. Differential mixing operation is shown in Fig.{~\ref{tmixer}} where subscripts P and N denote the original and inverted signals respectively.

\subsection{Two Qubit}
\begin{figure}[h]
\centering
\includegraphics[width=3in,height=3in]{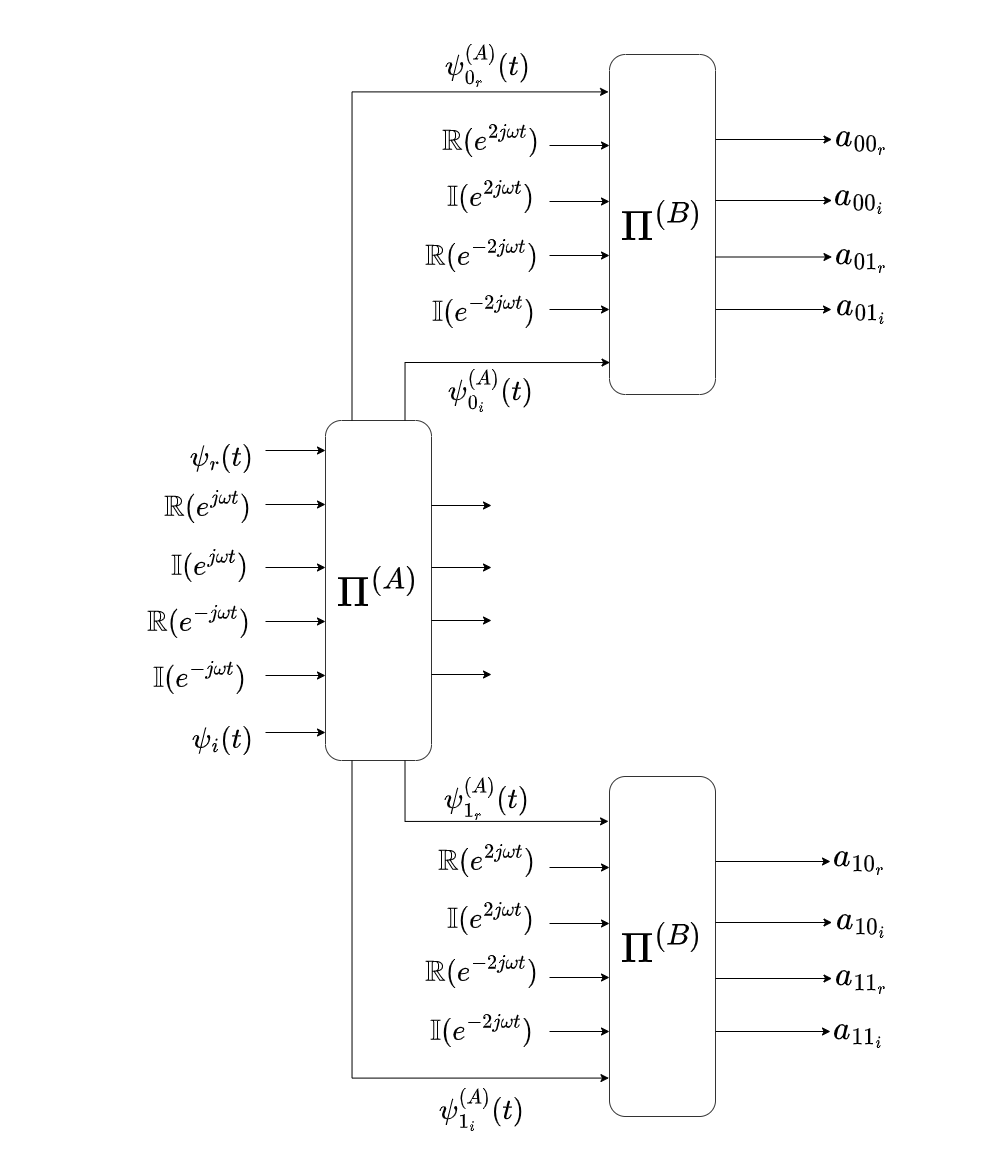}
\captionof{figure}{Recursive Projection}
\label{proj2}
\end{figure}

\begin{figure}[h]
\centering
\includegraphics[width=3.5in,height=2.2in]{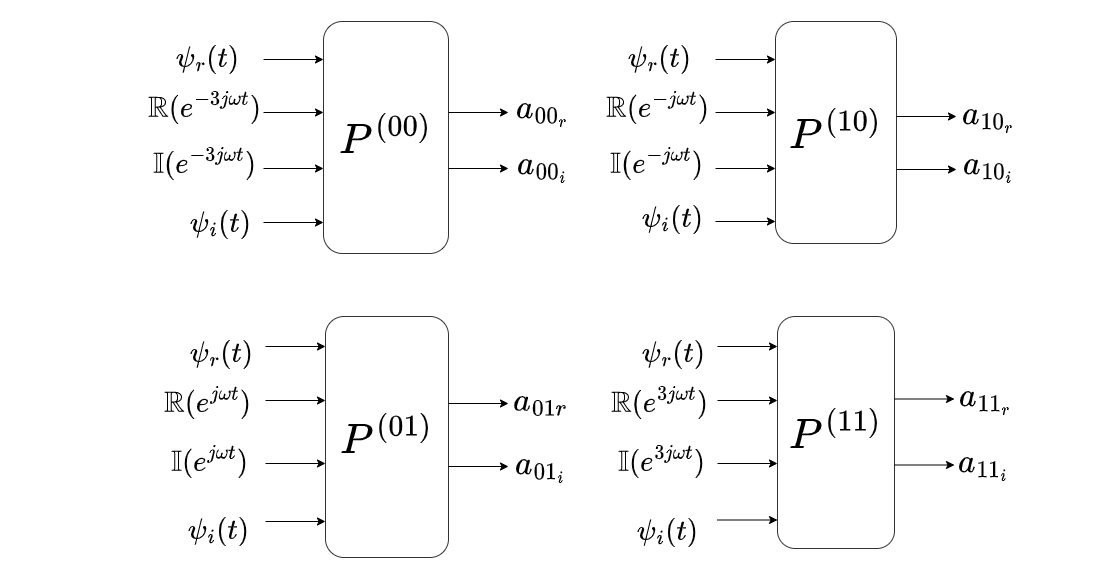}
\captionof{figure}{Parallel Projection}
\label{proj3}
\end{figure}

A two-qubit state comprising of two single qubits $A$ and $B$, corresponding to the base frequencies $\omega_{A} = \omega$ and $\omega_{B} = 2\omega$, can be represented by $\psi(t) = a_{00}e^{3j\omega t} + a_{01}e^{-j\omega t} + a_{10}e^{+j\omega t} + a_{11}e^{-3j\omega t}$ where $\psi(t) = \psi_{A}(t) \otimes \psi_{B}(t)$ and $\ket{0} = e^{3j\omega t}$, $\ket{1} = e^{-j\omega t}$, $\ket{2} = e^{j\omega t}$ and $\ket{3} = e^{-3j\omega t}$ are the four basis states. Projection of the state $\ket{\psi}$ on any of the basis can be achieved by performing the inner product in Equation 4 which is a multiplication followed by a filtering operation. Projection on $\ket{0}$ is as follows
\begin{equation}
    e^{-3j\omega t}\psi(t) = a_{00} + a_{01}e^{-4j\omega t} + a_{10}e^{-2j\omega t} + a_{11}e^{-6j\omega t}
    \label{decomp2}
\end{equation}
after filtering by a low pass filter with a cut-off frequency of $\omega$, we get 
\begin{equation}
     \braket{0|\psi} = a_{00}
\end{equation}
We require four projection blocks to obtain all the coefficients. The original paper suggests a recursive projection to do this with just 3 projections where we first project the state $\psi(t)$ onto the single qubit basis, say $\ket{0}_{A}$, by multiplying with $e^{-j\omega t}$ as follows
\begin{equation}
    e^{-j\omega t}\psi(t) = a_{00}e^{2j\omega t}  + a_{01}e^{-2j\omega t} + a_{10} + a_{11}e^{-4j\omega t}
\end{equation}
after using a bandpass filter with passband between $\omega$ and $3\omega$ we can filter out the terms where the first qubit is not zero (the last two terms in this case).
We then project the obtained signal onto the second qubit, say $\ket{0}_{B}$ as follows
\begin{equation}
    e^{-2j\omega t}(a_{00}e^{2j\omega t}  + a_{01}e^{-2j\omega t}) = a_{00}  + a_{01}e^{-2j\omega t}
\end{equation}
we then use a low pass filter to remove the higher frequency term. Thus, we obtained the coefficient of state $\ket{00}$ by first projecting it onto $\ket{0}_{A}$ and then onto $\ket{0}_{B}$. Similarly, we could obtain the coefficient of $\ket{01}$ by projecting $(a_{00}e^{2j\omega t}  + a_{01}e^{-2j\omega t})$ onto $\ket{1}$. This is demonstrated in Fig.{~\ref{proj2}} where $\Pi^{A}$ is the projection on qubit $A$.
Furthermore, the proposed parallel projection of Equation \eqref{decomp2} is shown in Fig.{~\ref{proj3}} where $P^{A}$ is the projection operation on a basis state rather than the entire qubit which makes it a single projection operation as in Fig.{~\ref{proj}}. 
The advantages of parallel projection over recursive projection are threefold. Firstly, the total number of projections got reduced to $4$ from $6$. Secondly, in the case of parallel projection, all the filters required are the same which are low pass filters with a cut-off frequency of $\omega$ whereas, we require two different types of filters out of which one is a low pass and the other is a bandpass. For the $n$ qubit system there would be $n$ different types of filters making the design more complicated. Thirdly, the proposed simplified mixing operation (Fig 6) would not be useful in the case of recursive projection as the harmonic rejection filter cannot be removed (which is a low pass filter with a cut-off frequency of $\omega_{x}+\omega_{y}$ where $\omega_{x}$ and $\omega_{y}$ are the frequencies of signals under multiplication) as the null space rejection filter after complex multiplication may not have an overlapping passband with the harmonic rejection filter. So, we cannot simplify the mixing operation. On the other hand, parallel projection makes all the null space rejection filters low pass with the same cut-off frequency ensuring an overlap with the harmonic rejection filter. There's one more advantage of using parallel projection which will be discussed in the following sections. Weighing in all the factors, we can see how parallel projection is simpler, faster, and consumes lesser area and power than recursive projection. \par
\begin{figure}[h]
\centering
\includegraphics[width=3.5in,height=5in]{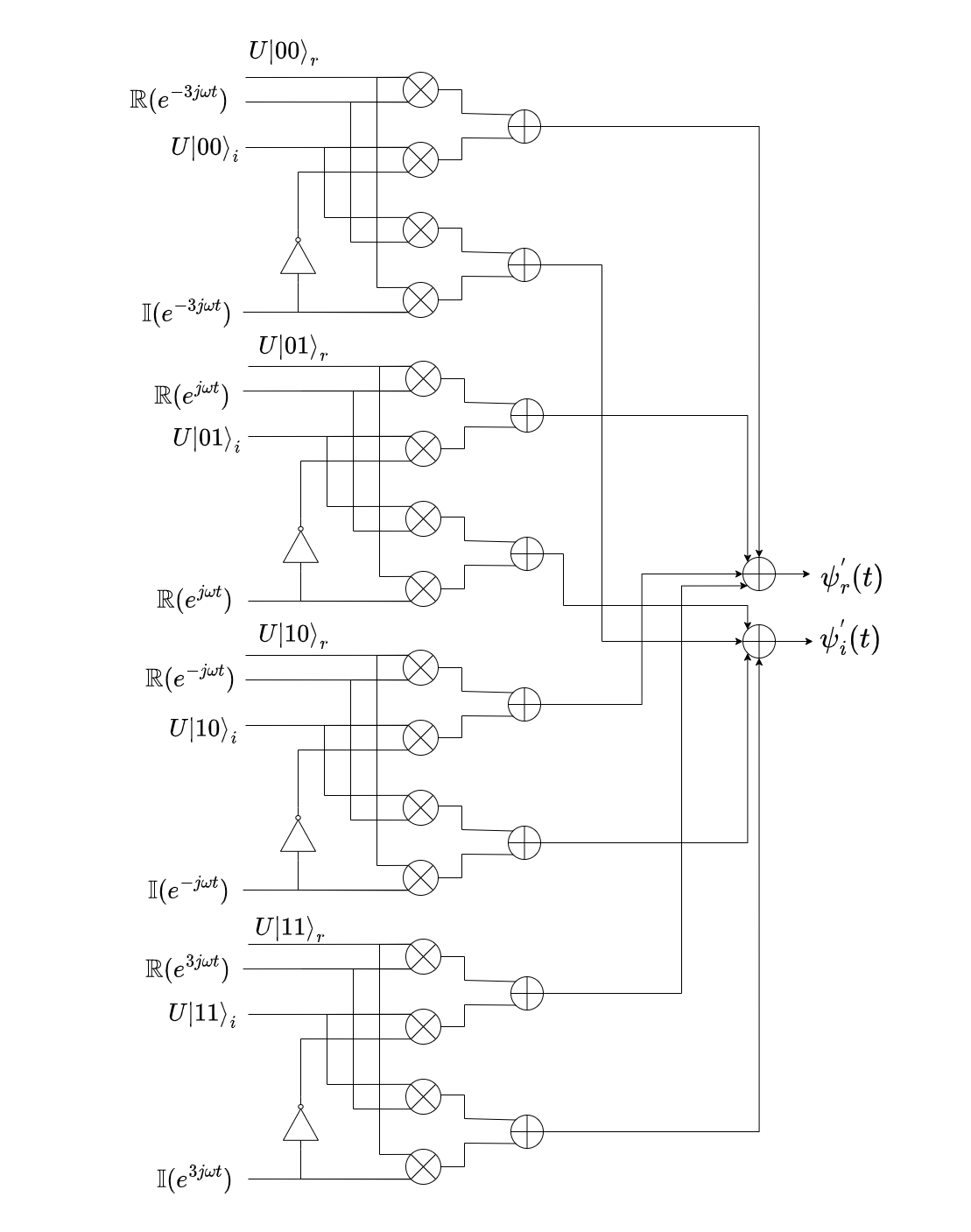}
\captionof{figure}{Recombination}
\label{combine}
\end{figure}

Finally, after projection, we apply the gate operation and combine all four coefficients each multiplied with its modified basis state as shown in Fig.{~\ref{combine}}.
\subsection{N Qubit}
Now that we established the design procedure for a two-qubit system, the design for an $n$ qubit one follows suit. Any $n$ qubit system can be represented as 
\begin{equation*}
    \ket{\psi} = \sum_{x=0}^{N}a_{x}\ket{\phi_{x}}
\end{equation*}
where, $N = 2^n$ and \par
\begin{equation*}
    \ket{\phi_{x}} = exp\bigg[\sum_{i=0}^{n-1}(-1)^x_{i}j\omega_{i}t\bigg]
\end{equation*}
We need $N$ basis signals and $N$ filters to perform the projection operation similar to Fig.{~\ref{proj3}}. After that, we will apply gate operation on the basis states and recombine them with the coefficients as shown in Fig.{~\ref{combine}}. 
\subsection{Measurement}
Once all the gate operations are done, we have to do the measurement which is nothing but finding the coefficients of all basis states which can be done by performing another projection operation on all basis states.

\section{Spatial Encoding}
The frequency-based encoding is limited by the number of qubits that can be represented in a given bandwidth due to the octave spacing scheme. To facilitate better scaling of the qubits the paper proposed a spatial encoding method. Consider a group of $M=2^m$ signals each representing a $N=2^n$ qubit state. This can be represented as
\begin{equation}
    \Psi(t) = \sum_{y=0}^{M-1}\psi_{y}(t)e_{y}
\end{equation}
where $e_{y}$ is a one hot vector with one in the row $y$ and zero elsewhere and
\begin{equation}
    \psi_{y}(t) = \sum_{x=0}^{N-1}a_{x,y}\phi_{x}(t)
\end{equation}
\begin{equation}
\Psi(t) = 
\begin{bmatrix}
    \psi_{0}(t)\\
    \psi_{1}(t)\\
    \vdots\\
    \psi_{M-1}(t)
\end{bmatrix}
\end{equation}
We define the inner product as
\begin{equation}
    \braket{\Psi'|\Psi} = \sum_{y=0}^{M-1}\braket{\psi'|\psi}
\end{equation}
This completes the Hilbert space description. Now we define how projection operation is defined for spatial qubits. We can decompose the state $\ket{\Psi}$ as follows
\begin{equation}
    \ket{\Psi} = \sum_{\bar{y_{i}}}\ket{\psi_{0,\bar{y_{i}}}}\ket{0,\bar{y_{i}}} + \sum_{\bar{y_{i}}}\ket{\psi_{1,\bar{y_{i}}}}\ket{1,\bar{y_{i}}}
    \label{sdecomp}
\end{equation}
This represents a projection of state $\ket{\Psi}$ on $\ket{e_{i}}$ where $\bar{y_{i}} = (y_{0}, y_{1}, \hdots, y_{i-1},y_{i+1}, \hdots y_{m-1})$ denote the binary representation of $y$ for all positions except the $i^{th}$ and the notation $y=(y_{i},\bar{y_{i}})$ denotes concatenation of the $i^{th}$ bit with the rest. Similarly for a two qubit projection on states $\ket{e_{i}}$ and $\ket{e_{j}}$ we have
\begin{equation}
\begin{split}
    \ket{\Psi} = \sum_{\bar{y_{ij}}}\ket{\psi_{0,0,\bar{y_{ij}}}}\ket{0,0,\bar{y_{ij}}} + \sum_{\bar{y_{ij}}}\ket{\psi_{0,1,\bar{y_{ij}}}}\ket{0,1,\bar{y_{ij}}}\\ + \sum_{\bar{y_{ij}}}\ket{\psi_{1,0,\bar{y_{ij}}}}\ket{1,0,\bar{y_{ij}}} + \sum_{\bar{y_{ij}}}\ket{\psi_{1,1,\bar{y_{ij}}}}\ket{1,1,\bar{y_{ij}}}
    \end{split}
\end{equation}
Now that we defined projection we will define the gate operation to complete the evolution of a quantum system. Suppose a unitary matrix $U$ is operating on the $i^{th}$ qubit
\begin{equation}
    U^{(i)}\ket{\Psi} = \sum_{\bar{y_{i}}}\ket{\psi_{0,\bar{y_{i}}}}\big(U^{(i)}\ket{0,\bar{y_{i}}}\big) + \sum_{\bar{y_{i}}}\ket{\psi_{1,\bar{y_{i}}}}\big(U^{(i)}\ket{1,\bar{y_{i}}}\big)
\end{equation}
where,
\begin{equation}
    U^{(i)}\ket{0,\bar{y_{i}}} = U^{(i)}_{00} \ket{0,\bar{y_{i}}} + U^{(i)}_{10} \ket{1,\bar{y_{i}}}
\end{equation}
\begin{equation}
    U^{(i)}\ket{1,\bar{y_{i}}} = U^{(i)}_{01} \ket{0,\bar{y_{i}}} + U^{(i)}_{11} \ket{1,\bar{y_{i}}}
\end{equation}
Substituting and rearranging gives
\begin{equation}
    \begin{split}
        U^{(i)}\ket{\Psi} = \sum_{\bar{y_{i}}}\bigg(U^{(i)}_{00}\ket{\psi_{0,\bar{y_{i}}}} + U^{(i)}_{01}\ket{\psi_{0,\bar{y_{i}}}}\bigg)\ket{0,\bar{y_{i}}} \\+ \sum_{\bar{y_{i}}}\bigg(U^{(i)}_{10}\ket{\psi_{1,\bar{y_{i}}}} + U^{(i)}_{11}\ket{\psi_{1,\bar{y_{i}}}}\bigg)\ket{1,\bar{y_{i}}}
    \end{split}
    \label{sadd}
\end{equation}
\subsection{Implementation}
To understand what Eq \eqref{sadd} represents, let's consider a two-spatial qubit system.
\begin{equation}
\ket{\Psi} = 
\begin{bmatrix}
    \psi_{0}(t)\\
    \psi_{1}(t)\\
    \psi_{2}(t)\\
    \psi_{3}(t)
\end{bmatrix}
\end{equation}
Consider a projection operation on the first qubit then Eq \eqref{sdecomp} becomes
\begin{equation}
\ket{\Psi} = 
\begin{bmatrix}
    \psi_{0}(t)\\
    \psi_{1}(t)\\
    0\\
    0
\end{bmatrix} + 
\begin{bmatrix}
    0\\
    0\\
    \psi_{2}(t)\\
    \psi_{3}(t)
\end{bmatrix}
\end{equation}
Now let's apply a gate $U$ on the first qubit.

\begin{equation}
U^{(0)}\ket{\Psi} = U^{(0)}
\begin{bmatrix}
    \psi_{0}(t)\\
    \psi_{1}(t)\\
    0\\
    0
\end{bmatrix} + U^{(0)}
\begin{bmatrix}
    0\\
    0\\
    \psi_{2}(t)\\
    \psi_{3}(t)
\end{bmatrix}
\end{equation}
this can be written as
\begin{equation}
\begin{split}
U^{(0)}\ket{\Psi} = U^{(0)}
\begin{bmatrix}
    \psi_{0}(t)\\
    0
\end{bmatrix} \otimes
\begin{bmatrix}
    1\\
    0
\end{bmatrix} + U^{(0)}
\begin{bmatrix}
    \psi_{1}(t)\\
    0
\end{bmatrix} \otimes
\begin{bmatrix}
    0\\
    1
\end{bmatrix} \\
+ U^{(0)}
\begin{bmatrix}
    0 \\
    \psi_{0}(t)
\end{bmatrix} \otimes
\begin{bmatrix}
    1\\
    0
\end{bmatrix} + U^{(0)}
\begin{bmatrix}
    0\\
    \psi_{1}(t)
\end{bmatrix} \otimes
\begin{bmatrix}
    0\\
    1
\end{bmatrix} 
\end{split}
\end{equation}
Simplifying this yields
\begin{equation}
U^{(0)}\ket{\Psi} = 
\begin{bmatrix}
    U^{(0)}_{00}\psi_{0}(t)+U^{(0)}_{01}\psi_{2}(t)\\
    U^{(0)}_{00}\psi_{1}(t)+U^{(0)}_{01}\psi_{3}(t)\\
    U^{(0)}_{10}\psi_{0}(t)+U^{(0)}_{11}\psi_{2}(t)\\
    U^{(0)}_{10}\psi_{1}(t)+U^{(0)}_{11}\psi_{3}(t)
    \label{smat}
\end{bmatrix}
\end{equation}
This is nothing but Eq \eqref{sadd} written in matrix form. Note that the summations in Eq \eqref{sadd} are just formal as the signals are spatially separated and summation just means combining them spatially. Can we achieve the same result without projecting the state? i.e., without separating the signals spatially. Consider a two-qubit gate operation $U\otimes I$ where $U$ acts on the first qubit and $I$ is an identity operator that acts on the second qubit. This can be represented by

\begin{equation}
\big(U\otimes I\big)\ket{\Psi} = \big(U\otimes I\big)
\begin{bmatrix}
    \psi_{0}(t)\\
    \psi_{1}(t)\\
    \psi_{2}(t)\\
    \psi_{3}(t)
\end{bmatrix}
\end{equation}

\begin{equation}
\big(U\otimes I\big)\ket{\Psi} = \Bigg(
\begin{bmatrix}
    U_{00} & U_{01}\\
    U_{10} & U_{11}\\
\end{bmatrix} \otimes
\begin{bmatrix}
    1 & 0\\
    0 & 1\\
\end{bmatrix} \Bigg)
\begin{bmatrix}
    \psi_{0}(t)\\
    \psi_{1}(t)\\
    \psi_{2}(t)\\
    \psi_{3}(t)
\end{bmatrix}
\end{equation}

\begin{equation}
\big(U\otimes I\big)\ket{\Psi} = 
\begin{bmatrix}
    U_{00} & 0 & U_{01} & 0\\
    0 & U_{00} & 0 & U_{01}\\
    U_{10} & 0 & U_{11} & 0\\
    0 & U_{10} & 0 & U_{11}\\
\end{bmatrix}
\begin{bmatrix}
    \psi_{0}(t)\\
    \psi_{1}(t)\\
    \psi_{2}(t)\\
    \psi_{3}(t)
\end{bmatrix}
\end{equation}
performing matrix multiplication gives
\begin{equation}
\big(U\otimes I\big)\ket{\Psi} = 
\begin{bmatrix}
    U_{00}\psi_{0}(t)+U_{01}\psi_{2}(t)\\
    U_{00}\psi_{1}(t)+U_{01}\psi_{3}(t)\\
    U_{10}\psi_{0}(t)+U_{11}\psi_{2}(t)\\
    U_{10}\psi_{1}(t)+U_{11}\psi_{3}(t)
    \label{matmul}
\end{bmatrix}
\end{equation}
which is the same as Eq \eqref{smat}. Hence, we conclude that spatial qubits can be implemented as matrix multiplications without ever needing to implement the projection operation. This equivalency extends to any $n$ qubit system. So for implementing $n$ qubits, we maintain $2^n$ spatial signals and operate them with matrices for state evolution. Here, we are omitting the case of projecting the state onto both frequency and spatial domains for a mixed gate operation as shown in the original paper \cite{main}. In the following section, we will compare the two encoding methods.
\subsection{Comparison with Spectral Encoding}
As in digital systems, here also we compare the number of floating point operations between the spectrally encoded method and spatially encoded method to capture the complexity of each. All the operations involved for each method and their requirement of multipliers and adders are listed in Table I. Note that $N=2^n$ where $n$ is the number of qubits realized.
\begin{table}[h!]
\caption{Floating Point Operations for various stages}
\centering
\begin{tabular}{ |c|c|c|c| } 
\hline
 Operation & Type & Multiplications & Additions\\
\hline
\multirow{3}{4em}{Projection} & Frequency-R & $8(N-1)$ & $4(N-1)$\\ 
& Frequency-P & $4N$ & $2N$  \\ 
& Spatial & - & -\\ 
\hline
\multirow{2}{4em}{\textit{n} Qubit Gate} & Frequency & $2N^2$ & $2N(N-1)$\\ 
& Spatial & $2N^2$ & $2N(N-1$) \\ 
\hline
\multirow{2}{4em}{Recomb} & Frequency & $4N$ & $2N+2(N-1)$ \\ 
& Spatial & - & - \\ 
\hline
\multirow{3}{4em}{Measure} & Frequency-R & $8(N-1)$ & $4(N-1)$\\ 
& Frequency-P & $4N$ & $2N$  \\ 
& Spatial & - & -\\ 
\hline
\multirow{3}{4em}{Total} & Frequency-R & $2N^2+20N-16$ & $2N^2+10N-10$ \\ 
& Frequency-P & $2N^2+12N$ & $2N^2+6N-2$ \\ 
& Spatial & $2N^2$ & $2N^2-2N$ \\ 
\hline
\end{tabular}
*Frequency-R - Frequency encoding with recursive projection.\\
*Frequency-P - Frequency encoding with parallel projection.\\
*Frequency - common for both
\end{table}

\begin{table}[h!]
\caption{Floating Point Operations for one gate operation}
\centering
\begin{tabular}{ |c|c|c|c| } 
\hline
 $n$ & Type & Multiplications & Additions\\
\hline
\multirow{3}{0.5em}{$1$} & Frequency-R & $32$ & $18$\\ 
& Frequency-P & $32$ & $18$  \\ 
& Spatial & $8$ & $4$\\ 
\hline
\multirow{3}{0.5em}{$2$} & Frequency-R & $96$ & $62$\\ 
& Frequency-P & $80$ & $54$  \\ 
& Spatial & $32$ & $24$\\  
\hline
\multirow{3}{0.5em}{$3$} & Frequency-R & $272$ & $198$\\ 
& Frequency-P & $224$ & $174$  \\ 
& Spatial & $128$ & $112$\\ 
\hline
\multirow{3}{0.5em}{$4$} & Frequency-R & $816$ & $662$\\ 
& Frequency-P & $704$ & $606$  \\ 
& Spatial & $512$ & $480$\\ 
\hline
\multirow{3}{0.5em}{$5$} & Frequency-R & $2672$ & $2358$ \\ 
& Frequency-P & $2432$ & $2238$ \\ 
& Spatial & $2048$ & $1984$ \\ 
\hline
\multirow{3}{0.5em}{$6$} & Frequency-R & $9456$ & $8822$ \\ 
& Frequency-P & $8960$ & $8574$ \\ 
& Spatial & $8192$ & $8094$ \\ 
\hline
\end{tabular}
\end{table}

Table II lists the number of multipliers and adders required for one gate operation qubit for various qubits. From this table, it is clear that the spatial encoding method is advantageous in terms of saving power and area. Also, note that spectrally encoded states require filters and comparators which are not needed for spatial encoded states. Recombination at the end of every gate operation is also not required for spatial states and lastly, measurement doesn't require any projections, it can be done by simply measuring the peak-to-peak amplitudes of the signals.\par
So far we have compared the methods considering them as two input digital floating point operations to understand the system-level complexity of the methods. Now we will do a circuit-level comparison based on the number of opamps each method consumes. Table III tabulates the number of opamps required for one gate operation split into stages for both spectral and spatial encoding. Similarly, Table IV summarizes the number of opamps per gate operation for various qubits. From these two tables, it is clear that even from a circuit-level view spatial encoding is exponentially more advantageous in terms of saving area and power when compared to any of the spectral encoding methods. As from Table III, we can see the difference between \textit{Spatial} and \textit{Frequency-P} is $8N$ which grows exponentially with $n$ (no. of qubits).
\begin{table}[h!]
\caption{No of Opamps for various stages}
\centering
\begin{tabular}{ |c|c|c|} 
\hline
 Operation & Type & Opamps\\
\hline
\multirow{3}{4em}{Projection} & Frequency-R & $8(N-1)$\\ 
& Frequency-P & $4N$\\ 
& Spatial & -\\ 
\hline
\multirow{2}{4em}{\textit{n} Qubit Gate} & Frequency & $4N$\\ 
& Spatial & $4N$\\ 
\hline
\multirow{2}{4em}{Recomb} & Frequency & $2N$\\ 
& Spatial & -\\ 
\hline
\multirow{3}{4em}{Measure} & Frequency-R & $8(N-1)$\\ 
& Frequency-P & $4N$\\ 
& Spatial & - \\ 
\hline
\multirow{3}{4em}{Total} & Frequency-R & $20N-16$\\ 
& Frequency-P & $12N$\\ 
& Spatial & $4N$\\ 
\hline
\end{tabular}
\end{table}

\begin{table}[h!]
\caption{Number of opamps for one gate operation}
\centering
\begin{tabular}{ |c|c|c| } 
\hline
 $n$ & Type & Opamps\\
\hline
\multirow{3}{0.5em}{$1$} & Frequency-R & $24$\\ 
& Frequency-P & $24$\\ 
& Spatial & $8$\\ 
\hline
\multirow{3}{0.5em}{$2$} & Frequency-R & $64$\\ 
& Frequency-P & $48$\\ 
& Spatial & $16$\\  
\hline
\multirow{3}{0.5em}{$3$} & Frequency-R & $144$\\ 
& Frequency-P & $96$  \\ 
& Spatial & $32$ \\ 
\hline
\multirow{3}{0.5em}{$4$} & Frequency-R & $304$ \\ 
& Frequency-P & $192$  \\ 
& Spatial & $64$\\ 
\hline
\multirow{3}{0.5em}{$5$} & Frequency-R & $624$  \\ 
& Frequency-P & $384$ \\ 
& Spatial & $128$\\ 
\hline
\multirow{3}{0.5em}{$6$} & Frequency-R & $1264$ \\ 
& Frequency-P & $768$ \\ 
& Spatial & $256$\\ 
\hline
\end{tabular}
\end{table}

Taking all these points into consideration, we choose the following configuration for implementing a $6$ qubit system. To exploit all the advantages of spatial qubits we choose $n=0$ spectral qubits encoded with $m=6$ spatial qubits. With this configuration, we implemented Grover's Search Algorithm and Quantum Fourier Transform as described in the following sections.

\section{Grover's Search Algorithm}
\begin{figure}[h]
\centering
\includegraphics[width=3.5in,height=2in]{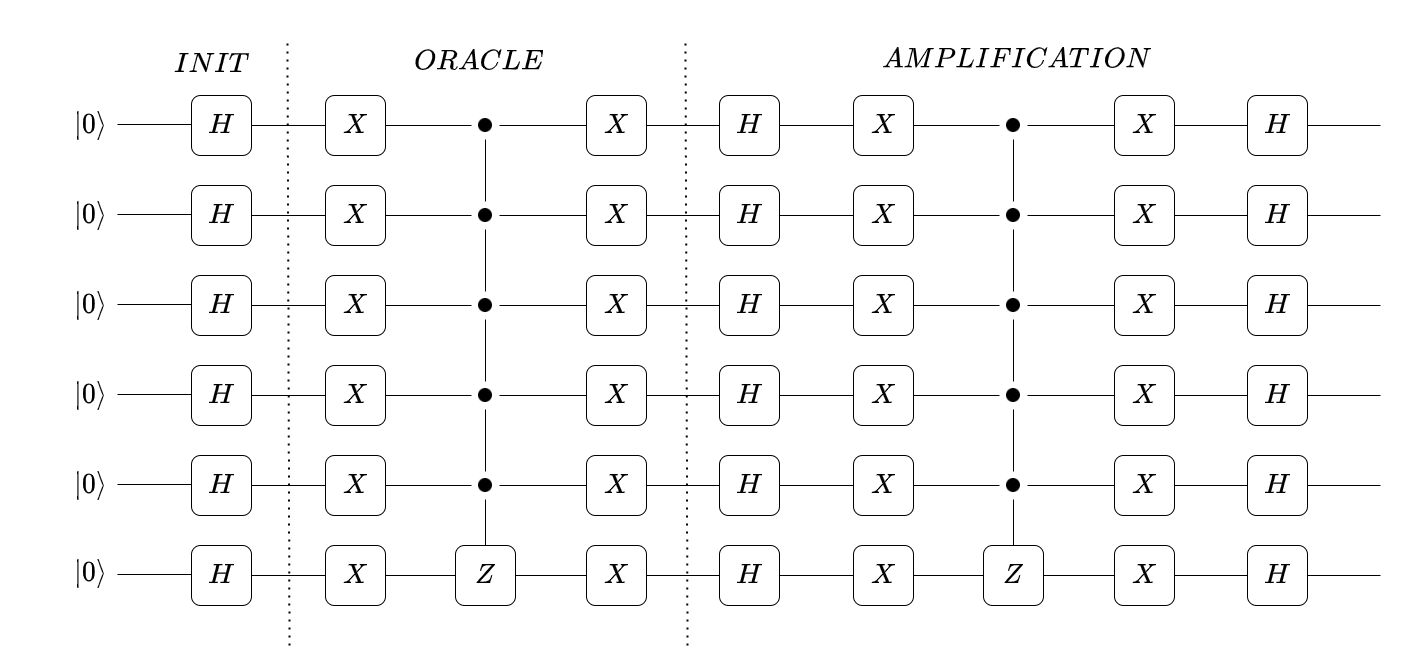}
\captionof{figure}{GSA Circuit}
\label{gsa}
\end{figure}

Grover's search algorithm for a six-qubit system is as shown in Fig.{~\ref{gsa}}. This consists of initialization, oracle, amplification, and measurement. The amplification block is applied multiple times to obtain the required accuracy. Measurement is not shown in the figure for simplicity.
There are a couple of things to note here. Firstly, the oracle is designed to search for the state $\ket{000000}$. To search for any other state a different oracle must be used. General oracle can be obtained by just removing the $X$ gate wherever the qubit is in $\ket{1}$ state. For example to search for the state $\ket{000001}$ we need to remove the $X$ gate from the first row and keep the remaining $X$ gates intact. Secondly, we don't need to decompose the six qubit gates into a universal set of gates ($C-NOT$ and 2 qubit gates) as we are dealing with analog signals and are not restricted by the decoherence or dephasing time or the fault tolerance of the system. Lastly, we need to implement only three types of gates here: a six-qubit Hadamard ($H$) gate, a six-qubit $X$ gate, and a $CCCCC-Z$ gate whose matrix forms are given as follows
\begin{equation}
H = 
\begin{bmatrix}
    1 & 1\\
    1 & -1\\
\end{bmatrix}
\end{equation}
then a six-qubit Hadamard gate becomes 
\begin{equation}
    H^{\otimes 6} = H\otimes H\otimes H\otimes H\otimes H\otimes H
\end{equation}
Note that $H$ is unnormalized here. Similarly for
\begin{equation}
X = 
\begin{bmatrix}
    0 & 1\\
    1 & 0\\
\end{bmatrix}
\end{equation}
\begin{equation}
    X^{\otimes 6} = X\otimes X\otimes X\otimes X\otimes X\otimes X
\end{equation}
Finally, the quintuple-controlled $Z$ gate can be written as 
\begin{equation}
    CCCCC-Z = I_{1}\otimes I^{\otimes 5} + I_{2} \otimes CCCC-Z
    \label{recur}
\end{equation}
where 
\begin{equation}
I_{1} = 
\begin{bmatrix}
    1 & 0\\
    0 & 0\\
\end{bmatrix}
, \quad
I_{2} = 
\begin{bmatrix}
    0 & 0\\
    0 & 1\\
\end{bmatrix}
\quad and \quad
I = 
\begin{bmatrix}
    1 & 0\\
    0 & 1\\
\end{bmatrix}
\end{equation}
and $CCCC-Z$ can be found using the same Eq \eqref{recur} recursively. Given all the matrix forms we discuss the implementation of the algorithm below.
\begin{figure*}[h]
\centering
\includegraphics[width=6.5in,height=2in]{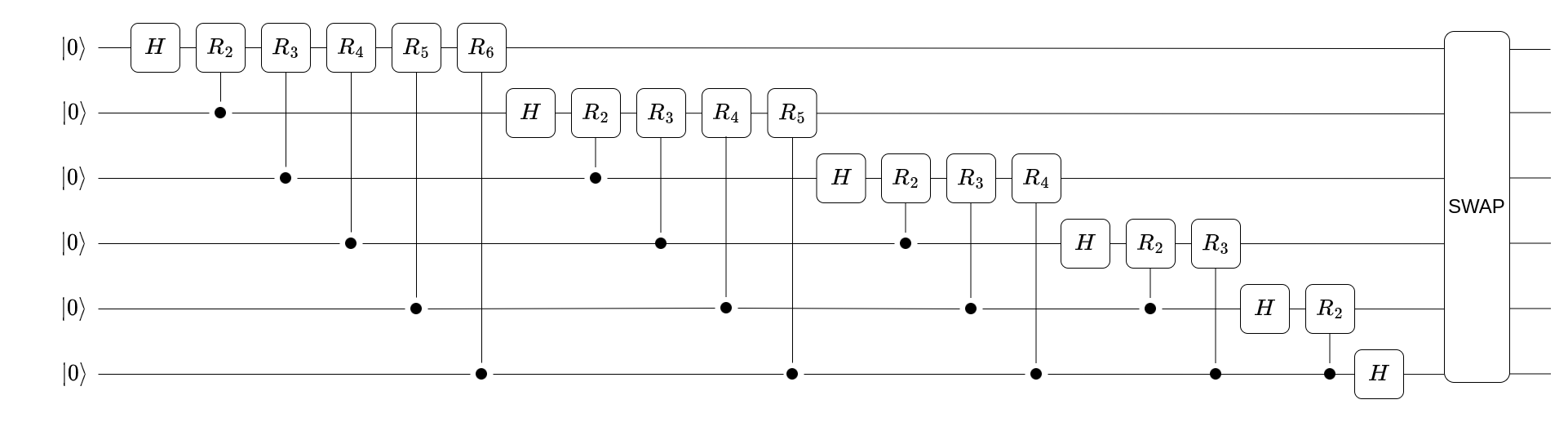}
\captionof{figure}{QFT Circuit}
\label{qft}
\end{figure*}
\subsection{Implementation}
Consider a six-qubit spatially encoded system with a zero spectrally encoded qubit.
\begin{equation}
\Psi(t) = 
\begin{bmatrix}
    \psi_{0}(t)\\
    \psi_{1}(t)\\
    \vdots\\
    \psi_{63}(t)
\end{bmatrix}
\end{equation}
where, $\psi_{x}(t) = a_{x}\phi(t)$. This means that all the spatial states have the same basic function $\phi(t)$ and they differ only by the amplitude of the signal. Now, the state $\Psi(t)$ can be written as
\begin{equation}
\Psi(t) = 
\begin{bmatrix}
    a_{0}\\
    a_{1}\\
    \vdots\\
    a_{63}
\end{bmatrix}
\psi(t)
\label{scale}
\end{equation}
and all the operations can be thought of as acting on the amplitudes instead of the entire state.
\subsubsection{Six Qubit H}
A Hadamard gate only consists of $1$ or $-1$ as elements of the matrix in its unnormalized form. Eq \eqref{matmul} suggests that if the elements of the unitary matrix are all $1$ or $-1$ we don't need to have multipliers. Suppose $H^{\otimes 6}\ket{\Psi} = \ket{\varPhi_{1}}$ then every row of $\ket{\varPhi_{1}}$ will be of the form $a_{1}\pm a_{2} \pm \hdots \pm a_{62} \pm a_{63}$ with varying signs. Also, we don't need inverters for obtaining the negative amplitudes ($-a_{x}$) as discussed earlier, we employed a differential design paradigm where the negative signals are readily available. With these simplifications, the Hadamard gate can be implemented by using adders alone. To obtain each row of $\ket{\varPhi_{1}}$ we need to perform $63$ additions so we used an opamp adder with 64 inputs to perform all the additions at once.

\subsubsection{Six Qubit X}
In matrix form X gate simply boils down to a swap gate that just reverses all the states. Suppose $X^{\otimes 6}\ket{\Psi} = \ket{\varPhi_2}$ then $\ket{\varPhi_{2}}$ will be of the form
\begin{equation}
\varPhi_{2}(t) = 
\begin{bmatrix}
    a_{63}\\
    a_{62}\\
    \vdots\\
    a_{0}
\end{bmatrix}
\psi(t)
\end{equation}
So this requires no multiplications or additions as simple wiring can perform the gate operation.
\subsubsection{CCCCC-Z}
The same goes for the $CCCCC-Z$ gate. This operates only on the states where all the first five states are $\ket{1}$ which is only the last two states. Suppose $CCCCC-Z\ket{\Psi} = \ket{\varPhi_3}$ then 

\begin{equation}
\varPhi_{3}(t) = 
\begin{bmatrix}
    a_{0}\\
    a_{1}\\
    \vdots\\
    a_{62}\\
    -a_{63}
\end{bmatrix}
\psi(t)
\end{equation}
So this gate also doesn't require any multipliers or adders. In the next section, we discuss Quantum Fourier Transform (QFT).
\section{Quantum Fourier Transform}
The circuit for Quantum Fourier Transform is shown in Fig.{~\ref{qft}}. It consists of a single qubit Hadamard gate and two qubit-controlled rotation gates. In the end, we reverse all the qubits i.e., $\ket{abcdef} \rightarrow \ket{fedcba}$ which is represented by the SWAP gate in Fig.{~\ref{qft}}. To implement this circuit, we need to find the matrix representation of each gate. The first Hadamard gate can be written as $H\otimes I^{\otimes 5}$ and the first controlled $R_{2}$ gate can be written as $I\otimes I_{1} \otimes I^{\otimes 4} + R_{2}\otimes I_{2} \otimes I^{\otimes 4}$. Note that $I_{1}$ occupies the second place in the first term as it is the control qubit and $R_{2}$ occupies the first place in the second term as it is the target qubit. The rest of the gates can be obtained similarly. Also
\begin{equation}
R_{n} = 
\begin{bmatrix}
    1 & 0\\
    0 & e^{j2\pi/2^n}\\
\end{bmatrix}
\end{equation}
With all the matrix forms at hand, we can simply implement the matrix operations using only opamps. 
\section{Results}
\subsection{Waveforms}The implementation of both the algorithms was described as matrix operations on a vector of complex coefficients but in fact, these coefficients are amplitudes of a wave function $\phi(t)$ as shown in Eq \eqref{scale}. 

\begin{figure}[h]
\centering
\includegraphics[width=3.5in,height=2.2in]{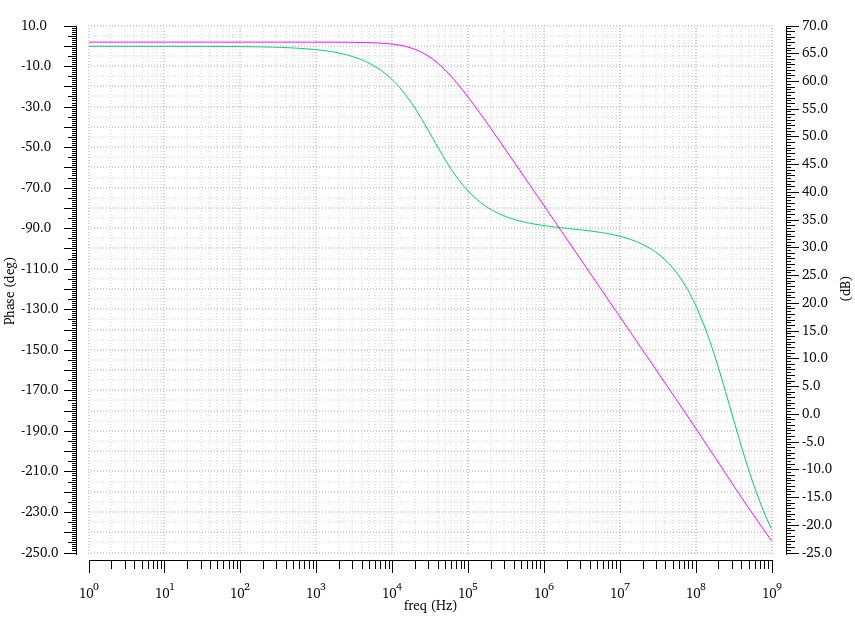}
\captionof{figure}{Opamp Characteristics}
\label{bode}
\end{figure}
For simulation purposes, we chose $\phi(t)$ to be a square wave of frequency $1KHz$. We designed a two-stage opamp with an open loop gain of $70dB$ and a gain bandwidth product of $1GHz$ and a phase margin of $70^\circ$. The bode plot of this opamp is shown in Fig.{~\ref{bode}}.

\begin{figure}[h]
\centering
\includegraphics[width=3.5in,height=2in]{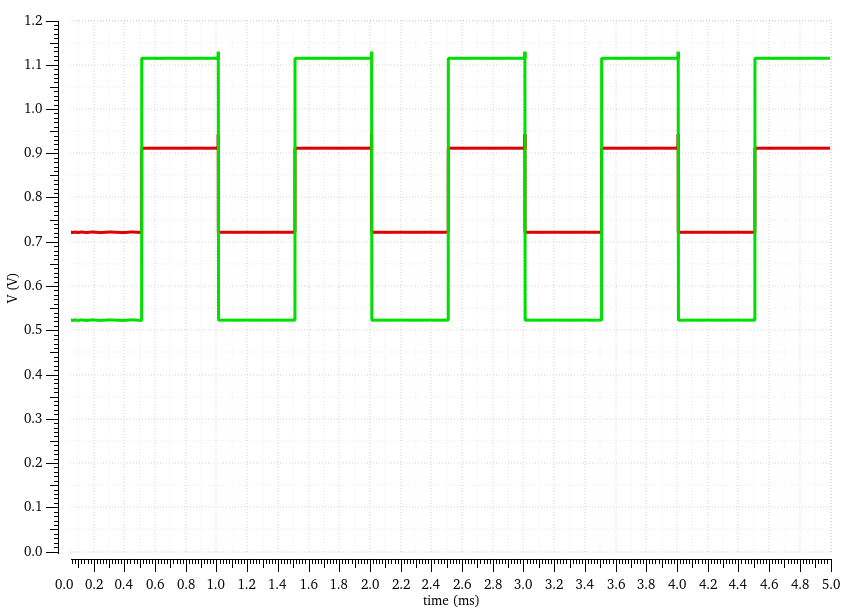}
\captionof{figure}{GSA Output}
\label{gsao}
\end{figure}

Using this opamp we implemented the matrix operations needed for both the algorithms in UMC 180nm process node. For Grover's algorithm, we got the following output in Fig.{~\ref{gsao}} when the input to the circuit in Fig 11 is $\ket{000000}$.

\begin{figure}[h]
\centering
\includegraphics[width=3.5in,height=2in]{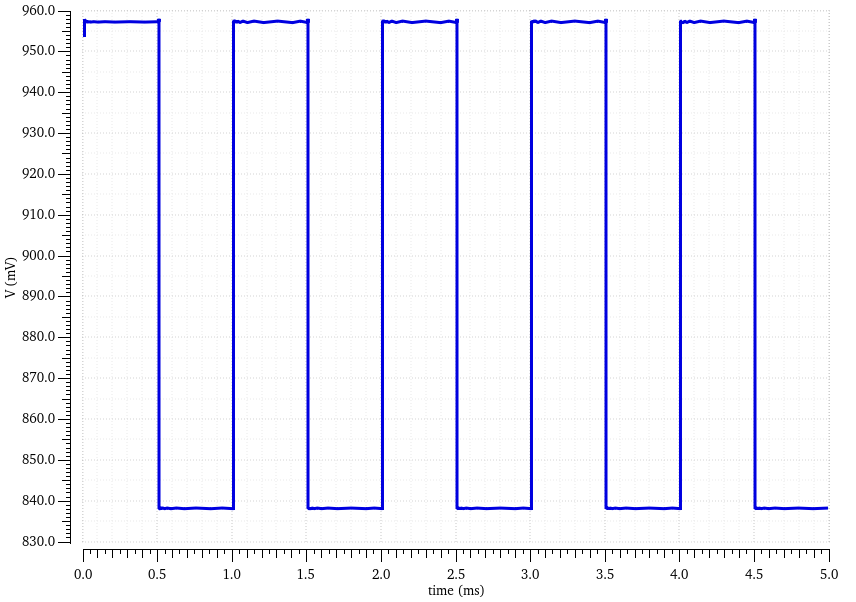}
\captionof{figure}{QFT Output}
\label{qfto}
\end{figure}

The green curve is the final state of $\ket{000000}$ whereas the red curve is the final state of $63$ other states. This is expected as the oracle was designed to search for the all-zero state and to amplify the amplitude of the same. We ran a MATLAB simulation to verify the results and found that the outputs are within $0.32\%$ of the expected outputs. Similarly, we got the following outputs from Quantum Fourier Transform in Fig.{~\ref{qfto}}.
In this case, all the states become equal after QFT when the input to the circuit in Fig.{~\ref{qft}} is $\ket{000000}$. We ran a MATLAB simulation for this too and verified that the expected outputs are the same as the obtained ones with almost $100\%$ accuracy. 

\subsection{Processing time} also calculated the processing time of both the algorithms and compared it to a Ryzen 5600x processor which has 6 physical cores running at 3.7 GHz of clock frequency. For GSA the time taken by AQED is about $1.056$ $\mu sec$ including all the propagation delays while the Ryzen took $1.669$ $\mu sec$ in the average case and $2.7$ $\mu sec$ in the worst case.
\subsection{Power Consumption}
Power consumption is an important factor when considering the performance of quantum computers because it is an established rule that the power required by a quantum computer scales exponentially with the number of qubits. So we ran the simulations to estimate the power consumed by both algorithms. GSA consumed a total power of $\textbf{151.292 mW}$ and QFT consumed a total of $\textbf{454.142mW}$. For comparison, Google's D-Wave 2X quantum computer uses about $\textbf{25kW}$ and almost all working quantum computers consumes in kilowatts. Most of the power is directed towards cooling the cryostat that enables the processor to sit at close to absolute zero which means a major chunk of power is not used for computation whereas an AQED uses the entire power for computation only. 
\subsection{Quantum Volume}
Quantum volume is a metric used to quantify the number of quantum circuits that can be processed on a quantum computer successfully \cite{QV}. The standard expression in the industry for quantum volume is as follows
\begin{equation}
    log_{2}V_{Q} = \argmax_{m \leq n} \{min[m,d(m)]\}
\end{equation}
where $n$ is the total number of qubits available and $d(m)$ is the circuit depth (i.e., number of gates that can be executed without decohering) when using only $m$ qubits out of $n$. \par
For AQED circuit depth is more than the number of qubits available as we emulated GSA and QFT which have more than $6$ gates. Hence, the quantum volume becomes
\begin{equation}
    log_{2} V_{Q} = n \Rightarrow V_{Q} = 2^n
\end{equation}
In our case $n=6$ so quantum volume is $V_{Q} = 64$ which is the best achievable quantum volume for any six-qubit quantum computer. IBM's Falcon r4 Montreal quantum computer accommodated 27 qubits to achieve a quantum volume of $64$ \cite{qv64}.
\section{Conclusion}
We have realized an Analog Quantum Emulation Device (AQED) which is intrinsically quantum and can take advantage of the inherent parallelism of quantum physics along with superposition. We first derived the system models of spectral encoding and spatial encoding methods separately and compared them against each other concerning complexity. On the other hand, we also simplified some aspects of spectral encoding such as projection. Later, we devised a $6$ qubit system emulating GSA and QFT using only opamps. We presented the results in the last section showing how an AQED can emulate an actual quantum computer.

\bibliography{reference}

\begin{thebibliography}{10}

\bibitem{feynman}
R.~P. Feynman, ``Simulating physics with computers,'' in {\em Feynman and
  computation}, pp.~133--153, CRC Press, 2018.

\bibitem{zoller}
B.~Zoller, ``Quantum information processing and communication,'' 2005.

\bibitem{google}
A.~Arute, ``Quantum supremacy using a programmable superconducting processor,''

\bibitem{super}
P.~Krantz, M.~Kjaergaard, F.~Yan, T.~P. Orlando, S.~Gustavsson, and W.~D.
  Oliver, ``A quantum engineer{\textquotesingle}s guide to superconducting
  qubits,'' {\em Applied Physics Reviews}, vol.~6, p.~021318, jun 2019.

\bibitem{trap}
J.~I. Cirac and P.~Zoller, ``Quantum computations with cold trapped ions,''
  {\em Physical review letters}, vol.~74, no.~20, p.~4091, 1995.

\bibitem{neutral}
G.~K. Brennen, C.~M. Caves, P.~S. Jessen, and I.~H. Deutsch, ``Quantum logic
  gates in optical lattices,'' {\em Physical Review Letters}, vol.~82,
  pp.~1060--1063, feb 1999.

\bibitem{main}
B.~R.~L. Cour and G.~E. Ott, ``Signal-based classical emulation of a universal
  quantum computer,'' {\em New Journal of Physics}, vol.~17, p.~053017, may
  2015.

\bibitem{adv}
B.~R. La~Cour, G.~E. Ott, and S.~A. Lanham, ``Using quantum emulation for
  advanced computation,'' in {\em 2017 IEEE Custom Integrated Circuits
  Conference (CICC)}, pp.~1--8, 2017.

\bibitem{parallel}
B.~R.~L. Cour, S.~A. Lanham, and C.~I. Ostrove, ``Parallel quantum computing
  emulation,'' in {\em 2018 {IEEE} International Conference on Rebooting
  Computing ({ICRC})}, {IEEE}, nov 2018.

\bibitem{acc}
A.~G. Blaiech, K.~B. Khalifa, C.~Valderrama, M.~A. Fernandes, and M.~H. Bedoui,
  ``A survey and taxonomy of fpga-based deep learning accelerators,'' {\em
  Journal of Systems Architecture}, vol.~98, pp.~331--345, 2019.

\bibitem{gsa}
L.~K. Grover, ``A fast quantum mechanical algorithm for database search,''
  1996.

\bibitem{qft}
D.~Camps, R.~V. Beeumen, and C.~Yang, ``Quantum fourier transform revisited,''
  {\em Numerical Linear Algebra with Applications}, vol.~28, sep 2020.

\bibitem{QV}
L.~Bishop, S.~Bravyi, A.~W. Cross, J.~M. Gambetta, J.~A. Smolin, and March,
  ``Quantum volume,'' 2017.

\bibitem{qv64}
P.~Jurcevic, A.~Javadi-Abhari, and Bishop, ``Demonstration of quantum volume 64
  on a superconducting quantum computing system,'' 2020.

\end{thebibliography}
\bibliographystyle{ieeetr}

\end{document}